\documentclass{aastex62}

\usepackage{amssymb}

\def\eg{{\it e.g.,} }
\def\etal{{\it et al. }}

\def\rs{r_{\rm S}}
\def\mbh{M_{\rm {bh}}}
\def\vr{v_{\rm r}}

\def\vp{v_{\rm \phi}}

\def\trp{\tau_{\rm r\phi}}

\def\omegk{\Omega_{\rm K}}
\def\alf{\alpha}
\def\lsim{\lower.5ex\hbox{$\; \buildrel < \over \sim \;$}}
\def\gsim{\lower.5ex\hbox{$\; \buildrel > \over \sim \;$}}
\def\simeq{\lower.3ex\hbox{$\; \buildrel \sim \over - \;$}}
\def\rt{r_{\rm t}}
\def\rout{\rm {OB}}
\def\as{a_{\rm s}}
\def\fadv{f_{\rm adv}}
\def\sun{M_\odot}
\def\fphi{f_{\rm\phi}}

\def\rin{r_{\rm in}}
\def\tt{\tilde{t}}
\def\geff{\gamma_{\rm eff}}
\def\fs{f_{\rm s}}

\def\neff{N_{\rm eff}}
\submitjournal{ApJ}
\shorttitle{Analyzing power index rule for ADAFs}
\shortauthors{Kumar \& Gu}
\begin{document}
\title{Self-similar solutions for finite size advection-dominated accretion flows}
\email{kumar@xmu.edu.cn, guwm@xmu.edu.cn}

\author{Rajiv Kumar}
\affil{Department of Astronomy, Xiamen University, Xiamen,
Fujian 361005, China}
\affil{ARIES, Manora Peak, Nainital, Uttarakhand 263002, India}

\author{Wei-Min Gu}
\affil{Department of Astronomy, Xiamen University, Xiamen,
Fujian 361005, China}

\begin{abstract}
We investigated effects on flow variables of transonic advection-dominated accretion flows (ADAFs) for different  
outer boundary locations (BLs) with a changing energy constant ($E$) of the flow. We used the ADAF solutions and 
investigated a general power index rule of a radial bulk velocity $(\vr\propto r^{-p})$ with different BLs, 
but the power index with radius for a rotation velocity and sound speed is unchanged. Here, $p\geq0.5$ is a power index. 
This power rule gives two types of self-similar solutions; first, when $p=0.5$ gives a self-similar solution of a first 
kind and exists for infinite length, which has already been discovered for the ADAFs by Narayan \& Yi, and second, when 
$p>0.5$ gives a self-similar solution of a second kind and exists for finite length, which corresponds to our new 
solutions for the ADAFs. By using this index rule in fluid equations, we found that the Mach number ($M$) and 
advection factor ($\fadv$) vary with the radius when $p>0.5$. The local energies of the ADAFs and the 
Keplerian disk are matched very well at the BLs. So, this theoretical study is supporting a two-zone 
configuration theory of the accretion disk, and we also discussed other possible hybrid 
disk geometries. The present study can have two main implications with a 
variation of the $p$; first, one that can help with the understanding of outflows and non-thermal spectrum 
variations in black hole candidates, and second, one that can help with solving partial differential equations for any 
sized advective disk. 
\end{abstract}
\keywords{accretion, accretion disks -- black hole physics -- hydrodynamics}
\section{Introduction} \label{sec:intro}
An accretion disk is powered by many galactic and extragalactic space objects, 
\eg compact objects, gamma-ray bursts, and young stellar objects. 
The compact objects are very compact and have a very high surface gravity compared to normal stars, 
\eg black holes (BHs), neutron stars, and white dwarfs.
In an accretion process, the accreting matter is compressed and becomes hotter, but may cool down by emitting 
radiation \citep{bbcf65,gfp03,fbg04}, and 
some part of the accreting matter may eject in the form of bipolar jets/outflows \citep{mrcpl92,jbl99,detal12}.  
These accreting sources also show spectral states variation in their activity cycles 
\citep{gfp03,fbg04}, \eg BH $X$-ray binaries (BHXBs).  
The BHXBs typically have two types of spectral states;  
one low/hard state (LS), which is radiatively inefficient and dominated by a power-law component 
with hard $X$-ray radiation, and second high/soft state (HS), which is radiatively efficient and 
dominated by multicolored blackbody radiation. 
These two states are connected with many intermediate hard/soft states \citep{fbg04}.
During the LS to intermediate/hard state, the BHXBs have weak or mildly relativistic to relativistic jets with 
quasi-periodic oscillations, but both features disappear during the HS.  
Interestingly, the disk's inner radius also decreases from LS to HS \citep{fbg04}.
These states and state transitions are thought to be manifested by the change in the disk size in the  
two-zone configuration theory of the accreting disks \citep{emn97,ds13}.

There are many accretion models that have explained many observed features of the black hole candidates (BHCs), 
\eg BHXBs and active galactic nuclei. 
The first popular and algebraic Keplerian disk (KD) model was given by \cite{ss73}. 
Which was basically nonadvective, optically thick, and had a cool flow,  
but it nicely explained the soft spectrum of the BHCs. 
Since this model was cool and ad hoc,   
therefore failed to explain other states of the BHCs. 
\cite{le74} also pointed out that the inner region of the disk was viscously and thermally unstable.
After this, many attempts were made to satisfy the inner boundary of the BH with advective nature of the flow, 
which can explain the nonthermal part of the spectrum and the outflows.
In the present study is also one of the attempts to fulfill the inner region of the KD 
with the help of ADAF solutions of different sizes. 
There are many hot advective models that have been developed for the accretion disk, 
\eg Thick tori \citep{pw80}, shock solutions \citep{c89}, and ADAFs \citep{ny94}. 
The shock solutions and ADAFs are generated from the same set of the fluid equations 
with different outer boundary conditions (BCs), which is explained on the basis of the energy constant of motion 
in this paper. 
All the advective accretion models stated above have different angular momentum distributions, 
and therefore the flows can have a single (inner or outer), or multiple critical points (CPs). 
The hot advective models are good at explaining the nonthermal part of the spectrum and the outflows. 
But intermediate states of the BHCs can be explained by the combination of the hot advective disk and the KD,  
making a hybrid disk \citep{emn97}. 
So, there are many studies made for the unification of the hybrid disks (or two-zone configuration theory) 
in order to make complete picture of observed states of the BHCs, 
like coupling of the ADAF and KD by \cite{h96,mk00,gpkc03,mm94,gl00,mlm00,llg04} 
and some authors with opposing opinions of this pair, \eg \cite{dt98,mgv01}.
Here, we believe in the unification of the hybrid model, 
although both kinds of the models are generated from different kinds of assumptions with fluid equations. 
Since the KD is a purely governed by algebraic equations, and used easily for the modeling of the soft spectrum 
of the BHCs, but is hard to combine with the inner transonic advective flow by fluid differential equations, 
we therefore used the technique of 
a self-similar solution approach of intermediate boundaries \citep{bz72}, which are obtained   
with the help of the transonic ADAF solutions. So together they can make the use of both models easier 
for the studies of the variation of the spectral states of the BHCs and outflows, by 
using the two-zone configuration accretion theory.

The self-similar solutions are qualitatively good with ADAFs \citep{ny94,nkh97}. 
They are used to
simplify partial differential equations (PDEs) into ordinary differential equations (ODEs) of the fluid 
for the studies of two-dimensional (2D) disk structures,  
with and without outflows, by many authors 
\citep{ny95a,xc97,BB99,BB04,XW05,xy08,gxll09,JW11,g12,g15,Jetl15} for HD flows and MHD flows 
\citep{sa16,mby16,has17,zmay18,gs18,bm18}, and also for the 
investigation of disk nonthermal radiation \citep{ny95b}. 
These authors have used the self-similar solution of the first kind, 
and all the three-velocities $\propto r^{0.5}$. 
The implications of self-similar solutions are not only limited to the study of accreting BHs, but also in cosmology, 
supernova, and other fields. 
The two-zone configuration theory 
can explain the various states of the BHCs with changing of the ADAF size. 
In the present study, we have defined general self-similar solutions (power index rule, $p$) of the second kind 
with the help of different sized ADAFs. 
So, these kinds of self-similar solutions can give an easier understanding of the two-zone radial 
theory, when one can reproduce results of \cite{ny95b,emn97},
and the outflows \citep{ny95a,JW11,zmay18} 
with changing only $p$, which changed the energy and transition radius of the flow.  
This study lacks two things; First, in the sense of quantitative ways because we have not considered explicitly relevant 
cooling in the fluid, and 
we will leave it for future studies. Second, the physics of transition from the KD to ADAF is not clear in this analytical study, but 
the local energy 
of the ADAF is nicely matched 
with the local energy 
of the KD at the transition radius 
as mentioned in section \ref{sec:anlys}. 
The aim of present study is to investigate the power-law index for various BLs of the ADAF,  
distinguish the hot and cool gas at the outer BL with nature of the general advective solutions on the basis of  
specific energy constant ($E$), and show the 
variation of the advective nature in the accretion disk with ADAF self-similar solutions. 
The structure of this paper is as follows: in next section 2 we talk about the fluid equations and assumptions, 
section 3 is for analyses, section 4 shows the results, and in the last section are the summary and discussions.

\section{The fluid Equations and Assumptions} \label{sec:eqns} 
We considered stationary advective viscous hydrodynamic fluid equations with axisymmetric in the spherical polar 
coordinates ($r, \theta, \phi$) around a nonrotating BH. 
For mathematical simplicity, we assumed Paczy\'nski-Wiita potential $\Phi=-G\mbh/(r-\rs)$\citep{pw80}, 
which represents the Schwarzschild geometry around the BH, 
where $\rs$ is the Schwarzschild radius. 
The fluid equations and flow variables are represented by geometrical units
and chosen as $2G=\mbh=c=1$ unless otherwise stated, where, $G$, $c$, and $\mbh$ are the universal
gravitational constant, the speed of light, and mass of the BH, respectively.
We then presented the accretion fluid equations of motion in the presence of viscosity 
along the equatorial plane.
The integrated form of the continuity equation gives the mass accretion rate ($\dot{M}$) of the flow with assuming 
a conical wedge flow disk 
under constant angle ($\theta$) from the axis of symmetry, and 
following \cite{kg18},
\begin{equation}
\dot{M}=-4\pi\rho \vr r^2cos\theta, 
 \label{ce.eq}
\end{equation}
the radial component of the Navier-Stokes equation,
\begin{equation}
 \vr \frac{d \vr}{d r}-\frac{\vp^2}{r}+\frac{1}{\rho}\frac{dP}{d r}+\frac{1}{2(r-1)^2}=0,
 \label{rc.eq}
\end{equation}
$\phi -$ component, 
\begin{equation}
 \vr \frac{d \vp}{d r}+\frac{\vp}{r}\vr=-\frac{1}{\rho r^3}\frac{d}{dr}(r^3\trp),
 \label{pc.eq}
\end{equation}
and the energy generation equation, 
\begin{equation}
 Q_{\rm adv}=\rho[\vr\frac{d\epsilon}{dr}-\frac{P}{\rho}\{\frac{\vr}{\rho}\frac{d\rho}{dr} \}]=Q^+-Q^-,
 \label{eg.eq}
\end{equation}
where $Q^+=\trp^2/{\eta}$ is a viscous heating rate and $Q^-$ is a cooling rate due to radiation.
Here $\trp=\eta(d\vp/dr-\vp/r)$ is the $r\phi$-component of the viscous stress tensor,
and $\eta=\alf P/\omegk$ is dynamical viscosity parameter. $\alf$ and $\omegk$ are 
the Shakura-Sunyaev viscosity parameter, and the Keplerian angular velocity, respectively.
$P$($=p_{\rm gas}+p_{\rm rad}$) is total pressure of the flow. $p_{\rm gas}={\rho\Theta}/{\tilde{t}}$ and 
$p_{\rm rad}$ are the gas pressure and radiation pressure, respectively. 
$\Theta=k_{\rm B}T/(m_{\rm e}c^2)$ is dimensionless temperature of the fluid, $\tilde{t}=\mu m_{\rm p}/m_{\rm e}$, 
where $\mu$, and $m_{\rm e}$ are the mean molecular weight of the gas, 
and mass of the electron, respectively. 
Here, we used two-component equation of state (EoS), where gas and thermal radiation both contributed to an energy 
density. Thus, total specific internal energy \citep{kfm08} is defined as
\begin{equation}
 \epsilon=\frac{p_{\rm gas}}{\rho(\gamma-1)}+3\frac{p_{\rm rad}}{\rho}=\frac{P}{\rho(\geff-1)},
 \label{eos.eq}
\end{equation}
where $\gamma$ is an adiabatic index, $N=1/(\gamma-1)$ is polytropic index, 
$\geff=1/[N\beta+3(1-\beta)]+1$ is effective $\gamma$, and $\beta=p_{\rm gas}/P$ 
is the gas pressure ratio. The sound speed is defined as
\begin{equation}
 \as=\sqrt{\frac{\geff P}{\rho}}.
 \label{sou.eq}
\end{equation}
Integrating Equation (\ref{rc.eq}) with respect to the radial distance ($r$) 
from the help of equations (\ref{eg.eq} and integrated form of equation \ref{pc.eq}) and after doing some algebra, 
we get constant of motion (${\cal E}$) and that is
\begin{equation}	
{\cal E}=\left[\frac{\vr^2}{2}+h-\frac{\lambda^2}{2r^2}+\frac{\lambda\lambda_0}{r^2}-\int q^-dr+\Phi\right], 
 \label{ec.eq}
\end{equation}
{where $\lambda=r\vp$ is a specific angular momentum of the flow and 
$\lambda_0$ is a specific angular momentum at the horizon.}
Since ${\cal E}$ comes from the first principle with the integration of all fluid equations for accretion flow, 
therefore it is a energy constant of motion 
in presence of any dissipation in the flow \citep{kc14}, which has information about the 
{angular momentum transportation energy, as well as rotational energy of the flow 
($-\lambda^2/{2r^2}+\lambda\lambda_0/r^2=-(\lambda-\lambda_0)^2/{2r^2}+\lambda_0^2/{2r^2}$),}
\deleted{lost rotational and}{energy loss ($-\int q^-dr$) due to} radiation 
{or other energy dissipations (if occurred)}\deleted{energies} during the journey of the fluid, 
and other local energies {($\vr^2/2, h, \Phi$)} of the flow. 
Here $q^-=Q^-/(\rho\vr)$ 
is a local radiative emissivity, and 
$h=\epsilon+P/\rho$ is a specific enthalpy of the flow. 
The nature of radiative emissivity is dependent on the optical depth of the flow, \eg 
$\tau\lesssim1$ nonthermal radiation, and $\tau>1$ thermalized radiation. 
If there is very weak or no radiative dissipation, assuming $Q^-\approx 0$ in the equation (\ref{ec.eq}) then  
the energy constant for 
radiatively inefficient accretion flow (RIAF) with following \cite{mgv01,gl04,kc13} is
\begin{equation}
{E}=\frac{\vr^2}{2}+h-\frac{\lambda^2}{2r^2}+\frac{\lambda\lambda_0}{r^2}+\Phi.
 \label{ec1.eq}
\end{equation}
This is also a constant of motion in presence of viscous dissipation only. 
Now, if we assume no viscosity in the flow and $\lambda=\lambda_0$, the 
equation (\ref{ec1.eq}) becomes,
\begin{equation}
 B=\frac{\vr^2}{2}+h+\frac{\lambda^2}{2r^2}+\Phi.
 \label{be.eq}
\end{equation}
This is a {local specific energy of the flow, and also known as the} canonical Bernoulli parameter, 
which is a constant of motion only for inviscid flow. 
Interestingly, all the three specific energies (${\cal E}, E$ and $B$) become equal at the BH horizon \citep{kc14}, 
because $(q^-, \trp)_{horizon}\rightarrow 0$ with $\lambda=\lambda_0$, 
and also a lowest local energy ($B$)\deleted{value} in the flow (Figure \ref{fig00}). 
{In other words, the energy constant of the motion is the lowest local energy of the flow, 
which has information about the energy loss and gain during the journey of the flow.}

The KD is cool, non-advective, rotation dominated ($\vr<<\vp$) and geometrical thin ($H/r<<1$) flow, 
thus we can use $\vr\sim0$ and $h\sim0$ 
in the equation (\ref{be.eq}) with $\lambda=\lambda_{\rm K}$, we get a local energy of the KD \citep{mgv01}, 
\begin{equation}
 B_{\rm K}\approx{\lambda_{\rm K}^2}/{2r^2}+\Phi=-1/4r, 
 \label{bk.eq}
\end{equation}
where $\Phi=-1/2r$ is used for the simplification of algebraic calculations instead of $\Phi=-1/2(r-1)$. 
\section{Analyses}\label{sec:anlys}
\begin{figure}[ht!]
\plotone{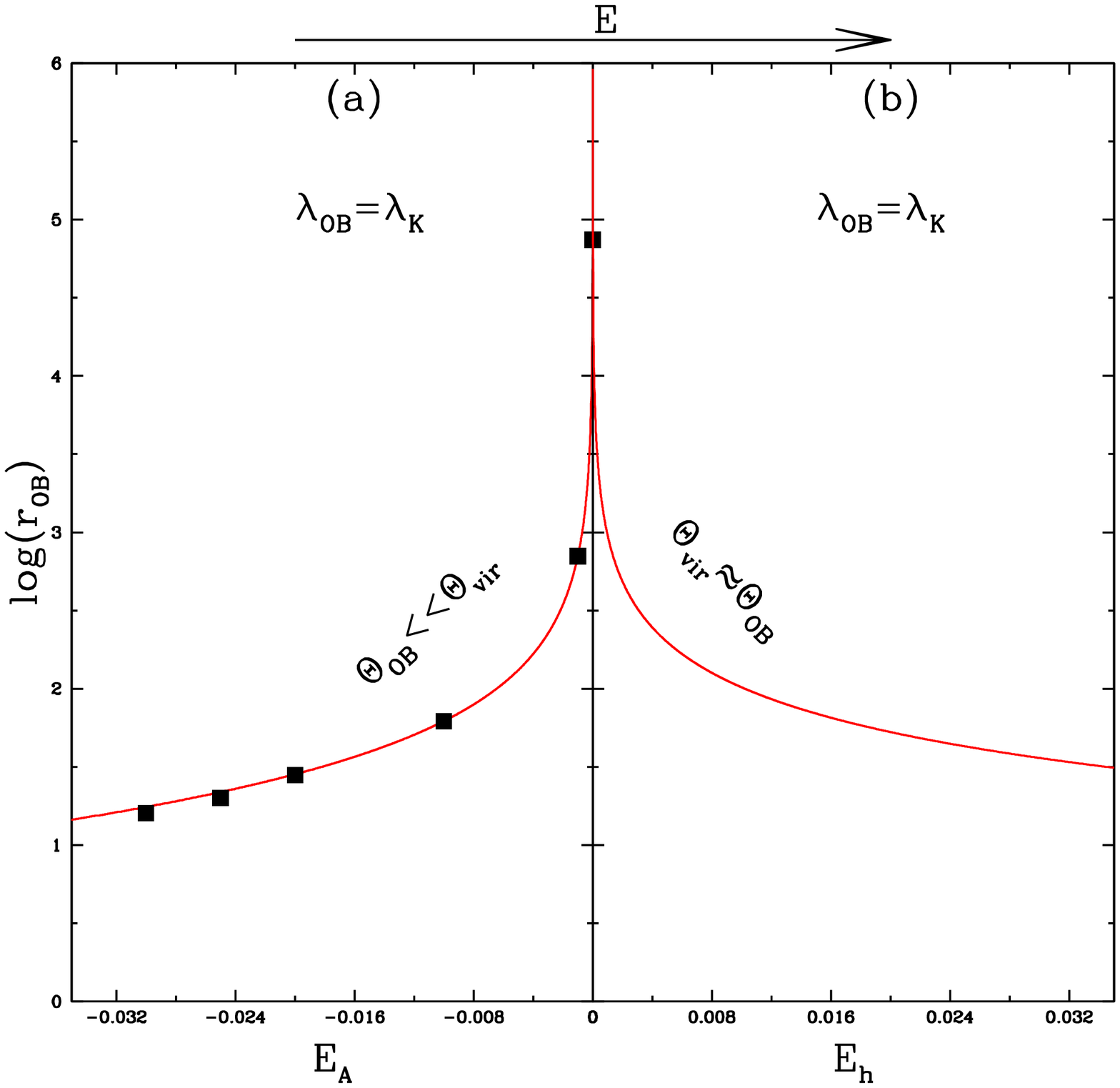}
\caption{Variation of the outer BLs ($r_{\rm OB}$) with $E$ (or $E_{\rm A}$ and $E_{\rm h}$). 
These BLs are corresponding only to transonic solutions with a single CP or three CPs. 
The 3-CPs are only found when $E>0$, and an outer and inner CP is connected with shock transitions \citep{kc13,lckhr16}.
The filled $\blacksquare$ points are represented in table \ref{tab:tab1} in the first and second columns. 
The arrow on top of the figure represents the direction of increasing $E$. $\Theta_{\rm vir}$ is the virial temperature.
\label{fig0}}
\end{figure}

We have the energy constants of motion (equations \ref{ec.eq}, \ref{ec1.eq}, and \ref{be.eq}), 
which come from the first principle with different assumptions in the fluid equations, 
\eg with and without viscosity or radiation. 
If we fixed the energy constant of motion then the outer BL of the transonic advective flow with 
corresponding nature of the flow is automatically fixed by satisfying accretion outer BCs 
\citep{kc13,kc14,ck16,kg18}. 
\deleted{means $\lambda_{\rout}\leq\lambda_{\rm K}$, $\vr\sim0$, and $\as\lesssim a_{\rm vir}$ 
when the energy constant is positive \citep{kc13,kc14}. 
Interestingly, when the energy constant is negative for the viscous flow then we get the ADAF solutions with 
satisfying the outer BCs 
\citep{nkh97,kg18}. 
Moreover, when the energy constant is positive then disk height of the flow at outer BL is always greater 
than the value of radial distance. 
If 
the energy constant is negative then the disk height of the global ADAF solution 
approaches to zero at outer BL \citep{nkh97,lgy99}, 
when the flow is in the vertical hydro-static equilibrium. 
Thus my notion is that the ADAF solutions can generated from some especial type of accreting gas (especially cool gas)
and location, \eg  Roche-lobe over flow through the Lagrangian point, and inner part of the KD.
We know that the local energy of the KD ($B_{\rm K}$) is also less than zero \citep{mgv01}, 
so there is a chance to generate ADAF solution. 
Therefore, we will compare the energies of these flows atleast at the BLs, and conclude the kind of possibilities
in this section.}We have investigated the outer BLs for the hot advective flows 
with the help of energy constant by using 
the BCs of the model solutions. 
{In other words, we have tried to connect the properties of the accreting gas at the outer BL with 
the kind of the advective solutions in this section, especially, gas temperature at the outer BLs.}
First, we are following the ADAFs' outer BCs by \cite{nkh97} with $\vr\sim0$, 
$h_{\rm OB}=\neff a_{\rout}^2<<a_{\rm vir}^2$ \deleted{$(\mbox{or}~a_{\rout}\sim10^{-3}r\omegk)$}({meaning 
the thermal energy should be negligible with the local rotational, and gravitational energies}), 
and $\lambda=\lambda_{\rm K}$ with assumed $q^-\rightarrow 0$ at $r=r_{\rout}$ in the equation (\ref{ec.eq}), 
giving us an energy constant at the BL,  
\begin{equation}
E_{\rm A}\approx \frac{\lambda_{\rm K}\lambda_0}{r_{\rout}^2}-\frac{\lambda_{\rm K}^2}{2r_{\rout}^2}+\Phi
=\frac{\lambda_0}{r_{\rout}\sqrt{2r_{\rout}}}-\frac{3}{4r_{\rout}}<0, 
 \label{eadafob.eq}
\end{equation}
where $\lambda_0\lesssim\lambda_{\rm m}$, and $\lambda_{\rm m}$ is the Keplerian angular momentum at an 
marginally stable orbit. 
We kept $\lambda_0$ in the $E_{\rm A}$ because $\lambda_0$ becomes comparable to $\lambda_{\rm K}$, 
when $r_{\rout}$ is close to the BH, otherwise $\lambda_0<<\lambda_{\rm K}$. 
The ADAF solutions with $E_{\rm A}$ are also obtained in \cite{kg18}. 
We calculated the size $r_{\rout}$ of the ADAFs corresponding to $E_{\rm A}$ for given $\lambda_0$ is represented in 
a panel (a) of Figure \ref{fig0}. 
Although $E_{\rm A}$ is a negative, but the flow is hot when it becomes sub-Keplerian and advective.  
The next type of outer BC is also rotating ($\lambda\approx\lambda_{\rm K}$) 
but hotter and temperature is approaching to the virial temperature at the outer BLs \citep{kfm08} 
with assuming $q^-\rightarrow0$. 
So $h_{\rm\rout}=\neff a_{\rm\rout}^2$, and $\vr\sim0$, 
where $a_{\rm\rout}\approx a_{\rm vir}=\sqrt{2\geff/3r}$ is a virial sound speed. 
Now, Equation (\ref{ec1.eq}) at the outer BL becomes
\begin{equation}
E_{\rm h}\approx h_{\rm\rout}+\frac{\lambda_{\rm K}\lambda_0}{r_{\rout}^2}-
\frac{\lambda_{\rm K}^2}{2r_{\rout}^2}+\Phi=
\frac{2\neff}{3r_{\rout}}+\frac{\lambda_0}{r_{\rout}\sqrt{2r_{\rout}}}-\frac{1}{12r_{\rout}}>0.
 \label{eadhot.eq}
\end{equation}
Here $E_{\rm h}>0$ gives the global advective solutions with/without shock solutions, which 
has been investigated in many studies \citep{bdl08,kc13,kc14,kc15,kscc13,kcm14,ck16}.
{In theoretical studies, we have found that the general advective solutions can be represented on 
a $E-\lambda_0$ plane \citep{kc13,kc14,kc17,ck16}. 
Here $E>0$ gives three CPs or single CP solutions, which depends on the energy  
and viscosity parameter of the flow, and $E<0$ gives two-CPs with close topology solution ($\alpha-$ type) 
when the flow is inviscid, and sub-Keplerian. 
For the viscous flow, $E<0$ also gives open topology global solutions, like, the ADAFs \citep{nkh97}.}
Now we want to categorize the outer BLs with a cool gas ($E<0$) and hot gas ($E>0$), because they both lead to 
different kinds of the accretion solutions. These BLs may depend on the properties of the accreting gases
and the BH feeding mechanisms,  
\eg Roche lobe overflow, star wind, accretion feedback, and inter-stellar gas etc., and  
a optical medium depends on the accretion rate. 
Moreover, the hot flow transfers more angular momentum than cool flow, for instance, 
ADAFs ($E<0$) have high $\lambda$ distribution than the shocked/smooth solutions with $E>0$.
{The accretion solutions with $E>0$ have the disk aspect ratio $H/r<1$ close to the BH 
($r\lesssim$ a few times a hundred), 
and $H/r\sim1$ far away from the BH \citep{lgy99,sd16}. 
The solutions with $E<0$ have the disk aspect ratio $H/r<1$ throughout the flow \citep{nkh97,lgy99}.}

The total energy of the KD at $r=\rin$, where the disk is chopped off, and 
$\lambda_0=\lambda=\lambda_{K}$ 
from equation (\ref{ec.eq}), becomes
\begin{equation}
 E_{\rm K}\approx \frac{\lambda_{\rm K}^2}{2\rin^2}-\int_{\rm in}^{\infty} q^-dr+\Phi=-\frac{1}{2\rin}, 
 \label{essd.eq}
\end{equation}
where $E_{\rm K}$ is a energy of the KD at the inner radius. 
$\int_{\rm in}^{\infty} q^-dr=1/4\rin$ is the energy loss due to radiation 
from $\infty$ to $\rin$ \citep{kfm08}. 
We now compare the energy of the different model solutions at the BLs, $E_{\rm h}>E_{\rm K}>E_{\rm A}$ 
and $(E_{\rm K}, E_{\rm A})<0$. 
Here, there is more chance that a high energy flow can convert to low energy flows with some dissipation. 
Now we can predict some possibilities for the hybrid disk geometries. First, all the three flows can coexist 
with a possible configuration, 
in that the outer part is the cool KD flow ($E_{\rm K}$)
and the inner part is the hot sub-Keplerian ADAF ($E_{\rm A}$) around the equatorial plane 
(like two-zone radial geometry), and 
both flows can be covered {partially or fully (depends on the smooth/shocked solutions and also
may depend on the sources of the hot accreting gas\deleted{since the inner part is also highly advective, 
and hot has almost same temperature distribution})} with hot sub-Keplerian gas ($E_{\rm h}$), 
because this flow has lowest 
$\lambda$ distribution {with higher the disk thickness from the flow with $E_{\rm A}$. For instance, 
\cite{kg18} have found that the supersonic and subsonic regions are formed above the equatorial plane in 
the inner part of some 2D disk structures, and both regions are connected with 
the shock like sharp transitions but close to the equatorial 
plane, the flow is always subsonic, before the inner CP}. 
Second\deleted{configuration}, the two flows can coexist {as described in \cite{wl91}}, 
\deleted{at the equatorial plane}and any other flow is negligible, 
\eg the outer part is the cool KD and the
inner is the hot sub-Keplerian ADAF, which is referred to as the two-zone radial configuration geometry flow \citep{emn97}, 
or the cool KD is covered with hot sub-Keplerian flow ($E_{\rm h}$) with 
extending to the BH horizon, which is referred to as the sandwich geometry or  two-component accretion flow (TCAF) \citep{ct95}. 
Third, only one flow can dominate and other flows have a negligible existence. 
\deleted{Thus, these hybrid disks configurations can possible from the hot and cool accreting gas sources at the BLs
with varying accretion rates.}

Here, the outer BCs with $E_{\rm h}$ and $E_{\rm A}$ give the sub-Keplerian transonic advective flows, 
but the KD is wholly subsonic, 
therefore there is an incomplete accretion solution. 
In order to make complete accretion for the KD onto the BHs then the flow needs to be transonic. 
By doing so there are two possibilities; first, the inner part of the KD can generate transonic advective 
solution with 
the help of internal instabilities \citep{le74,mm94}. 
Second, the gas of the inner part of the KD can evaporate due to external hot source \citep{mgv01}, 
like the hot flow above the equatorial plane, if it existed. 
Here, we assumed that there is no sufficient hot source is available, so the KD can generate transonic ADAF flow 
with some internal processes.\deleted{Therefore, we considered $E_{\rm K}$ is a initial energy of the flow 
and $E_{\rm A}$ is final energy of the flow, and 
energy difference is $\Delta E=E_{\rm initial}-E_{\rm final}=E_{\rm K}-E_{\rm A}=1/4\rin$ when $\rin=r_{\rout}$.}
{If we consider $E_{\rm K}$ is a energy at the inner radius of the KD and $E_{\rm A}$ is a energy of the hot flow 
then energy difference from the equations (\ref{eadafob.eq} and \ref{essd.eq}) is 
$\Delta E=E_{\rm K}-E_{\rm A}=[1/4\rin-\lambda_0/(2\rin^3)^{1/2}]>0$ at $\rin=r_{\rout}$ 
when $\lambda_0<\lambda_{\rm K}/2$, here $\lambda_{\rm K}$ at $\rin$. 
$\lambda_0$ depends on the viscosity and mass accretion rate of the flow.}
Here, $E_{\rm K}>E_{\rm A}$, 
so there is a possibility that {a part of the energy of} the KD can transfer into the advective flow 
and generate the ADAF solution, 
where $\vr$ and $h$ promptly become significant over rotations in the flow. 
\deleted{Interestingly, 
$\Delta E=1/4\rin$ is the same energy, which is lost due to radiation 
from the KD. 
In starting flow has $E_K$ energy but $\Delta E$ energy lost due to radiation during the KD then 
flow transit to ADAF with remaining $E_A$, and later}
$E_{\rm A}$ is distributed to the kinetic and 
thermal energy of the flow as the flow moves toward the BH. 
Now we can write $r_{\rout}=\rin=\rt$ in the rest of paper, where $\rt$ is a transition radius.
Apparently, this analysis looks almost perfect but needs further understanding of this transition process 
with addition of more physics. 
We are hoping for more exploration of this coupling in the future.
Moreover, the local energy of the KD and the ADAF is same at the $\rt$, 
which is $B_{\rm K}=B_{\rm A}=-1/4\rt$ from Equation (\ref{bk.eq}). 
{Here, $B_{\rm A}$ is a local energy of the ADAF at the outer BL, 
which also comes from the Equation (\ref{be.eq}), when the bulk velocity and thermal energy is negligible, 
meaning $v_r\sim 0$ and $h<<a_{\rm vir}^2$.}
So this energy analysis does not need any external heating\deleted{additional dissipation} at the transition radius, 
which is unlike to the statement of \cite{kn98}.
Moreover, ADAF solutions are the only solution to have a smooth variation from the Keplerian to sub-Keplerian $\lambda$ 
at the transition radius, and 
also has cool BL. 
If we compare equations (\ref{essd.eq}) and (\ref{eadhot.eq}) at $r_{\rout}=\rin$, $E_{\rm h}>E_{\rm K}$, 
here outer flow with $E_{\rm K}$ cannot transfer to transonic hot flow
unless there is some external heating at $\rt$ \citep{mgv01}. 
We know that the\deleted{From the outer BCs,} $H<<r$ for $E<0$ and $H\sim r$ for $E>0$ at the BL from 
the expression of the disc half-height, 
$H\sim\as/\omegk$. Thus, the ADAF solutions should generate from the cool and narrow space, and which matches 
with the inner region of the KD.
Therefore, we believed in the coupling of the KD and the ADAF, and
investigated the ADAFs with different sizes of the disk,  
which will certainly help in the understanding of the outflows variation with changing $\rt$, and 
it will be complementary to the 
two-zone configuration theory \citep{emn97}, when one can solve the 2D disk for the study of the outflows,  
as \cite{emn97} have studied only the variation of the spectral states with changing $\rt$.

In Figure \ref{fig0}, we plotted variations of BLs with energies of the advective flows for 
two types of outer BCs 
from the equations (\ref{eadafob.eq}) and (\ref{eadhot.eq}).
The variation of the outer BLs with both hot and cool gases are showing opposite behavior, with a variation of $E$.  
This means that high $E_{\rm A}$ has large BL 
and it decreases with decreasing $E$ in panel (a), but in panel (b), the BL is large for low $E_{\rm h}$, 
and it decreases with increasing $E$. Since BLs in the panel (a) are rotation dominated and in the panel (b) are thermally dominated 
but both have opposite behavior in the accretion flows. Means $\lambda_{\rm K}$ is high when the BL is larger, therefore $E_{\rm A}$ is high, 
and $\Theta_{\rm vir}$ is high when the BL is low therefore $E_{\rm h}$ is high.
Both kinds of gases can lead to transonic global solutions with large or shorter BLs. 
The interesting thing is that both have almost identical BLs, and 
the observed spectra of the BHXBs can be explained by the hybrid models (cool Keplerian and hot sub-Keplerian flows), 
which have different $\lambda$ distributions and advective nature of the flow. 
From our previous experince, the global shocked solutions occurred with large BL 
and hot gas at BLs means $0<E\lesssim0.007$ for $\alpha=0.01$ \citep{kc13}. 
And the ADAF disk solutions showed a comparatively large range of BLs and $E$ as shown in the table \ref{tab:tab1} and figure \ref{fig0}.
Both the energies have large as well as shorter BLs 
but in physical situations the BLs are large of the astrophysical objects. 
Therefore, we believed that the inner ADAF flow is originated from the outer KD, 
and together make the disk larger
with following the two-zone configuration theory. 
In next section, we will study the ADAF solutions for various outer BLs by solving fluid differential equations 
with following \citep{kg18}, and also 
find out the general self-similar solutions.
\section{Results}\label{sec:result}
In this section, we present the transonic ADAF solutions and 
estimated self-similar solutions of finite size. 
By using the self-similar solutions, we distinguished the advective and non-advective regions of the disk. 
Nonetheless, we got very interesting results that will be worthwhile in the understanding of 
the variation of the spectrum and jet states of the BHCs. 
This work is supporting previous studies \citep{mm94,ny95b,h96,emn97,kn98,gl00,mk00,mlm00,gpkc03,llg04}.
\subsection{Calculation of $p$ from ADAFs}
We used five disk flow parameters to investigate the transonic accretion solutions on the equatorial plane, 
which are ${E}, \lambda_0, \gamma, \beta$ and $\alpha$. 
Here, we investigated the ADAF solutions when $E<0$ with following outer BCs \cite{nkh97}. 
We integrated \deleted{the gradients of velocities and temperature equations after simplification of}the fluid 
differential equations (\ref{ce.eq}-\ref{eg.eq}) by four-order Runge-Kutta numerical method with
using same methodology for finding ADAF solutions as described in appendices of \cite{kg18}.

\begin{figure}[ht!]
\plotone{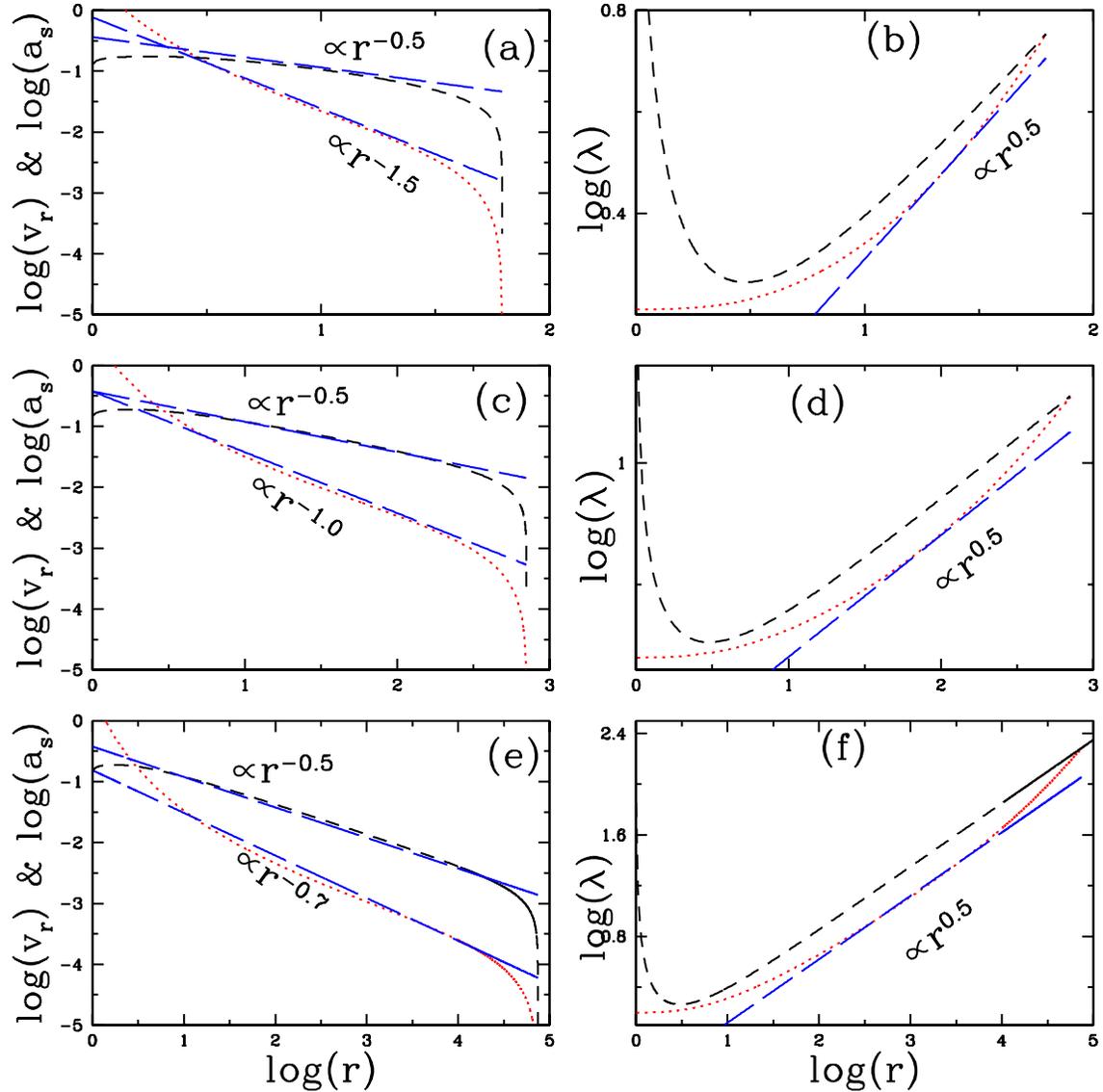}
\caption{Variations of the flow variables with radial distance ($r$) are plotted. 
We plotted variations of $\vr$, and $\as$ in the first column, and $\lambda$ in the second column. 
Panels (a, b), (c, d) and (e, f) are 
having different $E=-0.01, -0.001$, and $-0.00001$, 
respectively and other parameters, $\alpha=0.1, \beta=1.0, \gamma=4/3$ are fixed.
The curve long-dashed blue line represents the power law index in each panels, and 
dashed black line represents Keplerian distribution in the right column. 
\label{fig1}}
\end{figure}
We represented the ADAF solutions with different $E$, which gives us different size of the ADAF flow as 
shown in a Figure \ref{fig1}. 
Here we plotted bulk velocity ($\vr$, dotted red) and sound speed ($\as$, dashed black) curves together in a first column of the Figure \ref{fig1}, and 
the distribution of flow specific angular momentum ($\lambda$, dotted red) in a second column with the Keplerian distribution (dashed black). 
We scaled all the flow variables with radial power law 
which are represented by long-dashed blue line in the each panel of the figure. 
We got almost same power law variations for the $\as$ and $\lambda$, 
but different for $\vr$ with different BLs. 
The radial power law for $\as$ may vary when changing the space-time geometry of the central object.
The variation of the bulk velocity slope is obvious because the BCs of an inner (horizon, $\vr\rightarrow c$) 
and outer ($\vr\rightarrow 0$) of the BH accretion are fixed. 
Therefore, with variation of the BLs the flow velocity has to manage between these two BCs with adjusting slope of the bulk velocity.
The power index ($p$) of the $\vr$ are represented with $E$ corresponding their BLs in the table \ref{tab:tab1}. 
\begin{table}
\begin{tabular}{c c c c c c}
\hline \hline
Energy parameter & BLs & power index \\
${E}$ & $r_{\rm OB}$ & $p$ \\
\hline
$0>E>-1\times10^{-7}$&$>10^7$& 0.5 \\
\hline
$-1\times10^{-6}$&$\sim10^6$& 0.6 \\
\hline
$-0.00001$ & $\sim70,000$ & 0.7 \\
\hline
$-0.001$ & $\sim700$ & 1.0 \\
\hline
$-0.01$&$\sim62$&1.5 \\
 \hline
$-0.02$&$\sim28$&2.0 \\
\hline
$-0.025$&$\sim20$&2.1 \\
\hline
$-0.03$&$\sim16$&2.25 \\ 
\end{tabular}
\caption{We mentioned the outer BLs of ADAF disk with corresponding to the energy parameters ($E$), and
estimated $p$ from the velocity profile of the ADAF solutions.}
\label{tab:tab1}
\end{table}
The outer BLs of the transonic ADAFs with corresponding $E$ of table \ref{tab:tab1} are also represented 
on $E_{\rm A}$ vs $r_{\rm OB}$ plane 
with filled $\blacksquare$ points in the Figure \ref{fig0}. 
Interestingly, which are very well matched with directly calculated BLs from the equation (\ref{eadafob.eq}) 
in the Figure \ref{fig0}.
\subsection{Self-similar solutions of the ADAFs}\label{subsec:bimodel}
We analyzed global transonic ADAF solutions with various BLs and defined some general power rules of the flow variables 
with following table \ref{tab:tab1} and the Figure \ref{fig1},
\begin{equation}
 \vr \propto r^{-p}=-f_{\rm r}\alpha r^{-p},~~\as \propto r^{-0.5}=\fs r^{-0.5} ~~\mbox{and}~~\vp \propto r^{-0.5}=\fphi r^{-0.5},
 \label{powindx.eq}
\end{equation}
where $p\geq 1/2$. Here $p$ can also be written as $p=m+1/2$. 
Thus $r^m$ is a slope of the velocity curve, which depends on the ADAF disk size. 
When $p>0.5$ gives a self-similar solution of second kind, 
and exists for finite size length \citep{bz72}, which can not be derived from dimensional analysis. 
However, here we get them empirically and estimated from the ADAFs.
When $p=0.5$ then size is infinite (see Figure \ref{fig4} as $\rt\rightarrow\infty$ with $p=0.5$), 
so this gives a self-similar solution of first kind \citep{ny94}. 
Here, 
$(f_{\rm r},\fs,\fphi=\lambda/\lambda_{\rm K})\leq 1$ are scaling factors of the flow variables. 
The distribution of $\vr$ is much affected by $\alpha$, 
but other flow variables are not changed much by it, 
as seen in \cite{nkh97} and, more recently, \cite{kg18}. 
Therefore, we introduced $\alpha$ in the expression of $\vr$ and followed \cite{ny94}.
Interestingly, a variation of the Mach number ($M=\vr/\as\propto r^{0.5}/r^p$) with self-similar solutions is not 
a constant with the radius when $p>0.5$. 
So, this solved the problem of constant 
Mach number with self-similar ADAF solution ($p=0.5$), as mentioned by \citep{JW11,Jetl15}.

The outflows are very common from the hot advective disks, 
{which has been seen in the simulations \citep{nspk12,los13,snpz13,bywc13,bwy16,bygy16a,bygy16b,by19,ygnsbb15},  
and the observations \citep{ck12,wnm13,pkt18,mrlw19}}.
So, we assumed mass variation in the accretion disk due to the outflows and following \cite{BB99}, 
Equation (\ref{ce.eq}) becomes
\begin{equation}
-\dot{M}=4\pi\rho\vr r^2 cos\theta=\dot{M}_{\rm b}\left(\frac{r}{r_{\rm b}}\right)^s,
 \label{mloss.eq}
\end{equation}
where $s$ is the mass loss parameter \citep{kg18}. Here
$r_{\rm b}$ is an outer radius of the outflowing disk, and $\dot{M}_{\rm b}$ is the mass-accretion rate at $r_{\rm b}$.
The advection factor is defined by \cite{ny94,emn97}, which is
\begin{equation}
 \fadv=\frac{Q_{\rm adv}}{Q^+},
 \label{fdef.eq}
\end{equation}
where $Q_{\rm adv}$ and $Q^+$ are already mentioned in the equation (\ref{eg.eq}). 
We used equations (\ref{eos.eq}), (\ref{sou.eq}), (\ref{powindx.eq}) and (\ref{mloss.eq}) in (\ref{fdef.eq}), 
and after some simplifications we get
\begin{equation}
\fadv=\frac{4}{9}\frac{(\neff+p+s-2)f_{\rm r}}{\fphi^2 r^{(p-0.5)}}.
 \label{fadv.eq}
\end{equation}
Here $\fadv$ depends mainly on the $p$, therefore $r$, and also on $\fphi, f_{\rm r}$ and $\beta$. 
The $\fadv$ is independent of $\fs$ means independent of sound speed. 
Since variation of the sound speed 
is a mostly property of the central objects, 
means depends on the gravitational strength of the BHs \citep{kc17}. 
Although, $\dot{M}$ and $\beta$ are significantly changed the values of the temperature in the disk, which make hot or cool disk flows 
depending on the optical depth of the medium. 
But the total pressure is almost unchanged, 
and change in the gas pressure is compensated by radiation pressure, thus the sound speed is almost unchanged.

Now we would like to analyze equation (\ref{fadv.eq}) and calculate the $\fadv$ with $p$ values.
First, we assumed $p=1/2$ which also corresponds to the variation of the velocity like free-fall. 
So $\fadv$ is independent of the radial distance, and $\fadv=4(\neff+s-1.5)/9$ with $f_{\rm r}=\fphi=1$. 
Now it depends only on $\beta$ and $s$. 
The values of $\fadv$ for $\neff\mid_{max}=3~(\beta\rightarrow0)$ and $\neff\mid_{min}=1.5~(\beta\sim1)$ are 
$2/3$ and $0.0$ when $s=0$, respectively. 
The $\fadv$ is also found between these limits by \cite{g12,g15} 
with using the self-similar solution of the first kind.
Here $\fadv=0$ is only possible very far away from the BH with very cool gas. 
$\fadv<1$ means the viscous heating is more than the advective cooling, 
thus the flow becomes hotter and radiative in-efficient. 
So in order to balance the energy equation (\ref{eg.eq}), we need the outflows \citep{g12,g15} means $s>0$. 
Thus, we get $9/4\geq s\geq 3/4$, which depends on the $\beta$. Here, $s\sim3/4$
is consistent with many simulations \citep{omnm05,ywb12,ybw12,yyob14,Jetl15}, when $\neff\sim3$. 
For the self-similar solution of 
the second kind, 
when $p>1/2$ then $\fadv$ depends on the radius, which is explored in a following subsection.
%
\subsection{Calculation of advection factor and BLs of Bi-models}
\begin{figure}
 \plotone{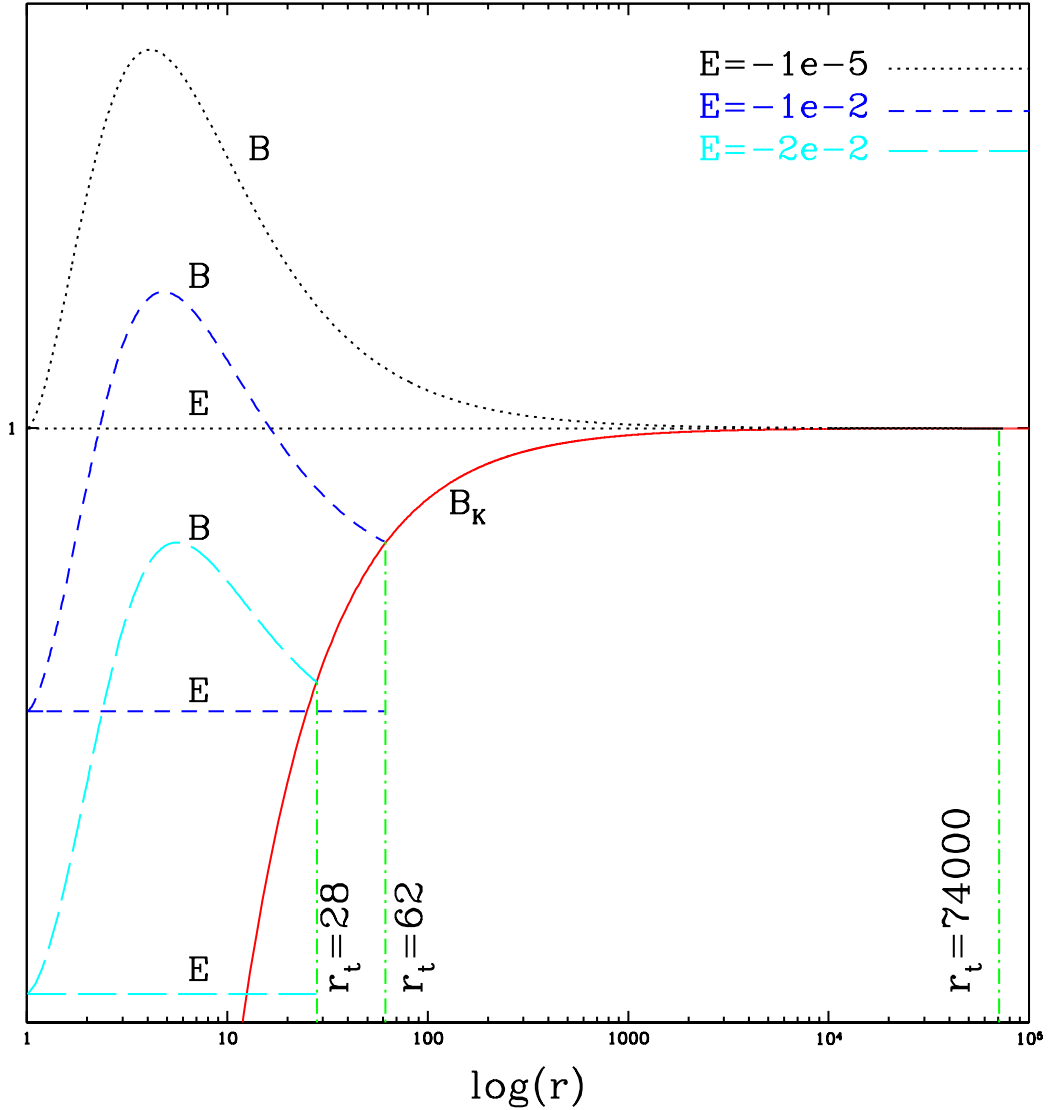}
 \caption{Variation of the local energies of the ADAFs ($B$) and the KD ($B_{\rm K}$) are plotted with 
 radial distance ($r$). The $B$ is plotted for three ADAF solutions as mentioned in the Figure (\ref{fig1}), 
 and corresponding three energy constants of motion ($E$) are also represented with same line style of each solution.  
 \label{fig00}}
\end{figure}
In Figure (\ref{fig00}), we have shown the variation of $B, E$ and $B_{\rm K}$ with $r$. 
Here we plotted $B$ and $E$ corresponding to three ADAF solutions, as shown in Figure (\ref{fig1}), 
where the values of $E$ are remain constant with the $r$ for each accretion solution. 
The outer BLs of the ADAF is represented with dotted-dashed line.
The $B$ is varying with the $r$ and has same value as $B_{\rm K}$ at the outer BLs of the ADAF or 
inner BLs of the KD, when $r=\rt$. Thus there is a possibility that both models can be connected.     
{Moreover, the local energy ($B$) of the flow becomes a lowest at the BH horizon, where $B=E$. 
So the constant of the motion of the flow ($E$) is a lowest local energy of the flow for a particular solution, 
which will be swallowed by the BH in the accretion flow.}

We look at the basic property of the both flows (ADAF and KD). 
The ADAF is the radiative inefficient means time scales of the radiative processes, which is much longer 
than the advection time but in the KD, this happens reversely, which makes the radiative efficient disk. 
Interestingly, the variation of $p$ changed the bulk velocity of the flow (equation \ref{powindx.eq}), 
thus the advection time.
\deleted{The self-similar solutions of second kind are applicable for finite size for advective flows, 
so we assume that other rest of part is applicable for the non-advective KD flow.}
\begin{figure}
 \plotone{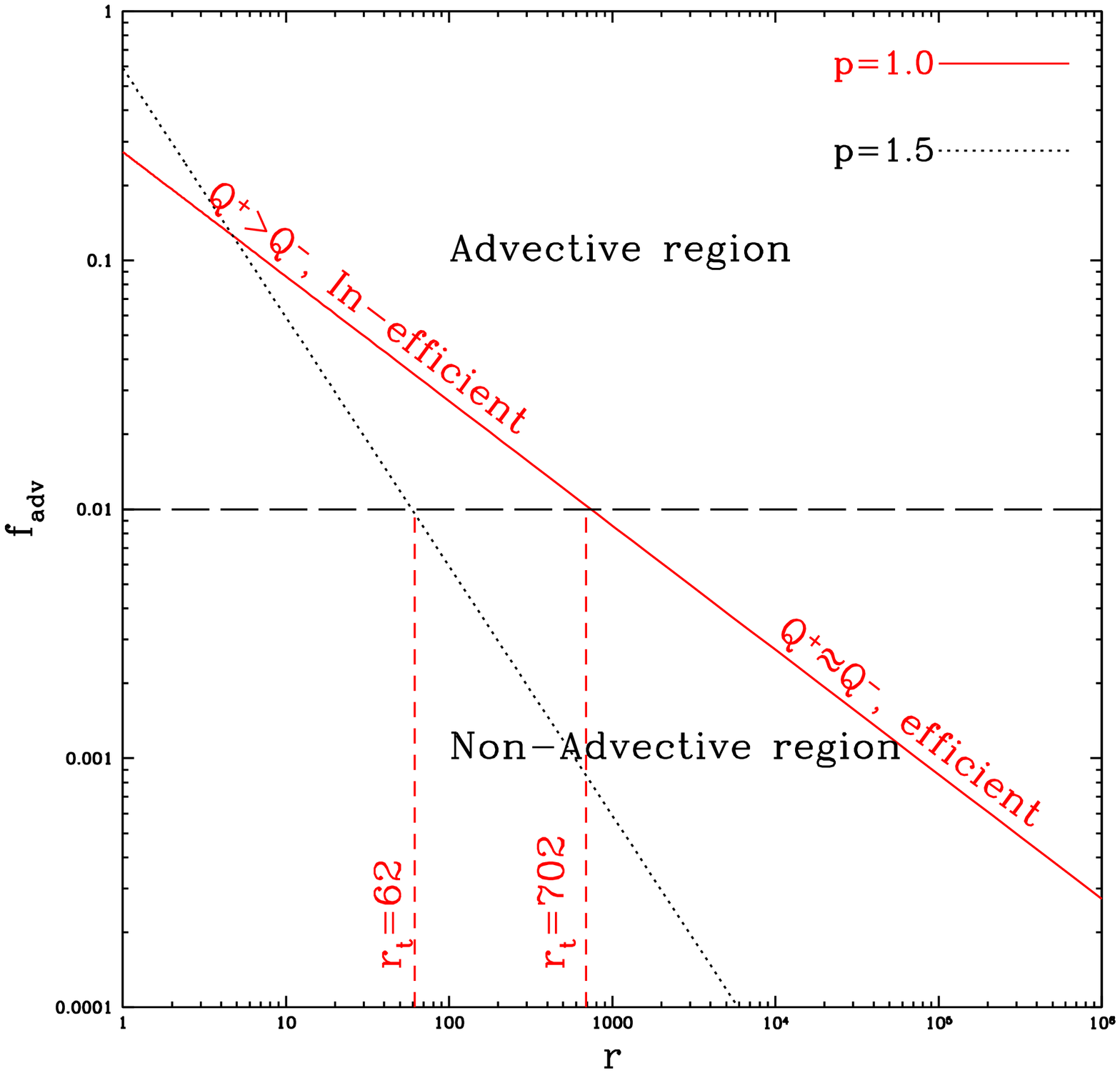}
 \caption{Variations of $\fadv$ with $r$ are represented with different $p=1.0$, and $1.5$. 
 A horizontal long-dashed (black) line is drawn for $\fadv=0.01$. 
A vertical dashed (red) lines are denoting transition radius of the flows with both the $p$.
 \label{fvsr}}
\end{figure}
The variation of $\fadv$ with $r$ is presented in a Figure (\ref{fvsr}) with different $p=1.0$ (solid red), 
and $1.5$ (dotted black), and 
fixed $\fphi=1$ and $s=0$. 
The vertical dashed (red) lines are representing transition radius ($\rt$) and the ADAF size. 
Since the KD has mass accretion rate so the disk must have a little advection, therefore 
we assumed $\fadv\lesssim 0.01$ with $\fphi=1$. 
Thus we can safely assume that the viscous heating rate ($Q^+$) is equal to radiative cooling rate ($Q^-$). 
In doing so, we have drawn a horizontal long-dashed (black) line in the Figure with $\fadv=0.01$, 
which separates the advective and non-advective regions and cuts both curves at the respective $\rt$. 
For a particular curve with corresponding $p$, the flow is advective and radiatively inefficient when $r<\rt$, 
and is non-advective and radiatively efficient when $r>\rt$. 
For $r<\rt$, we can also change the angular momentum distribution from Keplerian to sub-Keplerian values, 
which will also increase $\fadv$. 
In both curves, the advective region decreases with increasing $p$ but $\fadv$ arises faster 
and becomes higher at some location.  
Thus, my notion is that the outflow region definitely decreased with increasing $p$ because the ADAF size decreased,  
but we are not clear about a behavior of the outflows strength. 
Nonetheless, the advective factor is increasing faster with 
decreasing ADAF size (Figure \ref{fvsr}), so the outflow strength may increase with advection \citep{JW11}. 
We will leave this issue for a future study and communicate it as a separate paper. 

\begin{figure}
 \plotone{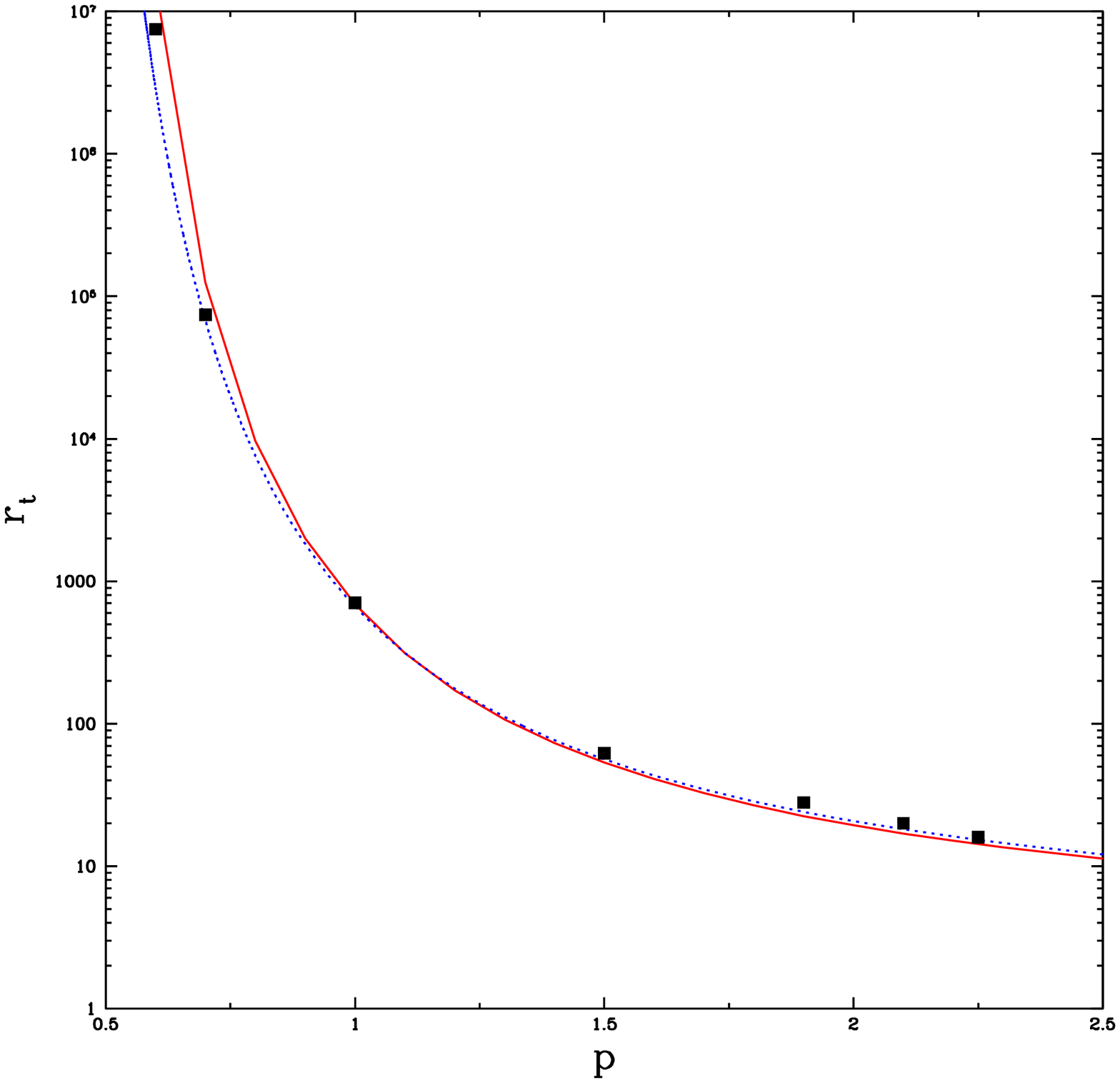}
 \caption{Variation of the transition radius ($\rt$) with $p$ are presented with the second and third columns from table \ref{tab:tab1} ($\blacksquare$ points),
 and calculated inner radius of the KD with assumed $\fadv\approx0.01$ (solid red line). The dotted line represents fitting curve.
 \label{fig4}}
\end{figure}
In Figure \ref{fig4}, we have represented variations of the BLs with $p$. 
A dashed blue curve is a fitting curve of the data points of the second and third columns of the table \ref{tab:tab1}, 
which is also a projection of $E_{\rm A}$ vs $r_{\rout}$ of the Figure \ref{fig0}a. 
This gives us the relation between the $p$ and\deleted{information about the $p$ to corresponding value of} the $\rt$, 
so the fitting formula is $p=a/log(b\rt)+c$, where $a=3.952, b=0.4614$ and $c=0.3215$. 
This fitting formula will be useful when one can study the effects on the outflows and emitted non-thermal 
radiation from the ADAFs with changing $\rt$.  
A solid red curve is corresponding to the inner radius of the non-advective flows 
or transition radius of the two flows, which is calculated from equation (\ref{fadv.eq}) when $\fadv=0.01$ with changing $p$. 
Here both curves are almost same. 
Hence the equation (\ref{fadv.eq}) is consistent with the BLs of the actual ADAFs and transition radius with 
assumed $\fadv\approx0.01$.
\section{SUMMARY AND DISCUSSION}
This study shows two main important points of using the self-similar approach;  
first, we can solve the PDEs and study the 2D disk structure with/without the outflows, 
and second, a study of the spectral states variations of the BHCs with the analytical approach. 
Although the numerical simulations better represent the fluid dynamics around the astrophysical objects, 
but the present study is useful in the understanding of the nature of the accretion solutions with the different 
outer BLs and outer BCs with hot or cool gases, 
Which will help in the studies of the accretion dynamics with variations of the emitted radiation and the outflows 
by the simulations.  
Here, we investigated the self-similar solutions of the second kind for the ADAFs, 
and we proposed the variations of the advective and non-advective region by using the self-similar solutions. 
%

In this study, we have explored transonic advective solutions with changing $E<0$. 
Interestingly, they have the single sonic point and easily fit with a single power law of the radius for a 
large range of the distances (Figure \ref{fig1}).
Therefore, we investigated the power index of the flow variable with different BLs. 
These power indices of the variables with finite BLs are identified as the self-similar solutions of 
the second kind \citep{bz72}. 
The first time, we identified this kind of solution for the ADAF disk around the BHs, 
as our best knowledge of this kind of accretion study. 
Although, the self-similar solution of the first kind was identified for the ADAF by \cite{ny94}.
In the accretion flow studies, many authors have only used the first kind solution ($p=0.5$), 
which gives a constant Mach number in the accretion flows but the second kind ($p>0.5$) gives radially 
dependent Mach number ($M\propto r^{(0.5-p)}$). Which is more realistic in the accretion flow. 
We used general self-similar solutions and solved the fluid differential equations and 
calculated the advection parameter ($\fadv$) with the radius (Figure \ref{fvsr}).
For the first kind solution ($p=0.5$), we get the advection parameter and the Mach number are independent of 
the radius, which is not reasonable in the accretion flow. 
Now, one can use the second kind solutions ($p>0.5$) in solving ODEs or PDFs for the accretion and outflows, 
which will give more better results, and one can also conclude that how the outflows are changing with the BLs. 
So, this will give the better understanding of the disk structure and the outflow generations. 

As we are interested in the two-zone configuration theory of the disk, which can explain spectral states 
variations of the BHCs with variations of the inner ADAF disk and the outer KD \citep{emn97}. 
Therefore, we investigated the possible coupling of 
the KD and ADAFs solutions on the basis of the advective nature (Figure \ref{fvsr}) 
and the energy of the flows (Figure \ref{fig00}).
As we know, the KD was constructed with the purely algebraic equations and non-advective in nature. 
Now the ADAFs also become in the algebraic in their form by obtaining the self-similar solution of finite 
size boundaries ($\rt$). 
So we assumed the KD is present when $r>\rt$ and interestingly, 
we found a very little advection at the $\rt$ and beyond, $\fadv\lesssim0.01$ 
for any value of the $p>0.5$ (Figure \ref{fvsr}). 
Thus 
we can safely apply other assumptions of the KD, which will give thermal BB emissivity.   
When $r<\rt$ is applicable for the ADAF self-similar solution of the second kind, 
and one can easily use them to calculate non-thermal radiation 
and the outflows like \cite{ny95a,ny95b} with different size of the ADAFs.
There is one more advantage of this study for the possible coupling of the bi-modals, we 
did not require any external source because $B_{\rm K}=B_{\rm A}$ when $E<0$, which is unlike to the conclusion of 
the previous studies \cite{kn98,mlm00,mgv01,llg04} when $E<0$. 
For ADAF Self-similar flow when $p=0.5$, we did not get the non-advective or cooling-dominated flow, and 
which is consistent with the conclusion of \cite{kn98}.
If we want to couple the KD with the advective sub-Keplerian flow from $E>0$ then 
we need an external hot source to evaporate the inner part of the KD \citep{kn98,mgv01}, because $E_{\rm h}>E_{\rm K}$. 

{Moreover, we have also predicted the different kinds of the hybrid disk geometries for the accretion flows 
depending on the temperature of the accreting gases at the BL. For instance, when the gas comes from large 
radii in two-phases (cold clumps embedded in hot diluted gas). 
In this case, the hybrid disk can be formed depending on the accretion rates. 
Here we considered the hot accretion rate ($E>0$) and cool accretion rate ($E<0$).
As the hot gas is very efficient 
in the angular momentum transportation at the large radii when the gas is captured by the disk. 
Therefore, the hot gas becomes quickly advective and can generate the smooth solution with single CP or the 
shocked solution with multiple CPs \citep{kc13,kc14,ck16}. 
On the other hand, the cold gas can settle in the form of the KD or 
the ADAF disk or both flows (the hybrid disk) around 
the equatorial plane with depending on the cool mode accretion rate \citep{emn97}. 
So, the maximum three kinds of the flow can coexist in the hybrid disk depending on the hot and cool accretion rates, 
and the viscosity. If the flow components are less than three then other flow(s) is/are negligible,  
which may configure as the two-zone radial flow geometry (inner hot ADAF and outer the KD) or like 
sandwich geometry (one flow or two-zone radial flows around the equatorial plane, covered with 
the other hotter shocked/smooth flow).
These kinds of geometries, with at least one hot flow, can help in the explanation of the outflows and 
emission of the very high energy photons due to the inverse comptonization of the 
soft photons of the KD with the hot post-shock region or inner part of the smooth advective flow.}

The timing and quantitative analysis 
of the BH feeding mechanisms with the variation of the 
spectral states, and the nature of the accreting gas sources 
are needed to explore more by the observational and theoretical studies. 
Since the flow with the cool gas ($E<0$) and the hot gas ($E>0$) at the outer BLs generate different 
kinds of the accretion solutions and $\lambda$ distributions.
These two points can help in the understanding of all the observed properties of the BHCs with 
changing the mass accretion rate and the viscosity. 
The final conclusion of our study is that
the both kind of the BLs (hot $E>0$, and cool $E<0$) may depend on the BH feeding mechanisms 
and the nature of the accreting gas sources (hot or cool).
{So, the qualitative and quantitative studies of the BH feeding mechanisms can help in the 
understanding of the BH accretion physics.}
\acknowledgments
 
We thank the anonymous referee for his helpful comments and suggestions, 
which improved our manuscript.
This work was supported by the National Natural Science Foundation of China under grant 11573023.
R.K. also thanks the IUCAA, Pune, India for the local hospitality and stay during the revising this manuscript.


\begin{thebibliography}{}
\bibitem[Barenblatt \& Zol'dovich (1972)]{bz72} Barenblatt, G. I., \& Zel'dovich, Y. B. 1972, AnRFM, 4, 285
\bibitem[Becker \etal (2008)]{bdl08} Becker, P. A., Das, S., \& Le, T. 2008,\apj, 677, L93
\bibitem[Blandford \& Begelman (1999)]{BB99} Blandford, R. D., \& Begelman, M. C. 1999, \mnras, 303L, 1
\bibitem[Blandford \& Begelman (2004)]{BB04} Blandford, R. D., \& Begelman, M. C. 2004, \mnras, 349, 68
\bibitem[Bowyer \etal (1965)]{bbcf65} Bowyer, S., Byram, E. T., Chubb, T. A., \& Friedman, H. 1965, Sci, 147, 394
\bibitem[Bu \etal (2013)]{bywc13} Bu, D.-F., Yuan, F., Wu, M., \& Cuadra, J. 2013, \mnras, 434, 1692
\bibitem[Bu \etal (2016a)]{bygy16a} Bu, D.-F., Yuan, F., Gan, Z.-M., Yang, X.-H. 2016, \apj, 818, 83
\bibitem[Bu \etal (2016b)]{bygy16b} Bu, D.-F., Yuan, F., Gan, Z.-M., Yang, X.-H. 2016, \apj, 823, 90
\bibitem[Bu \etal (2016)]{bwy16} Bu, D.-F., Wu, M.-C., \& Yuan, Y.-F. 2016, \mnras, 459, 746
\bibitem[Bu \& Mosallanezhad (2018)]{bm18} Bu, D.-F., \& Mosallanezhad, A. 2018, \aap, 615, 35
\bibitem[Bu \& Yang (2019)]{by19} Bu, D.-F., \& Yang, X.-H. 2019, \mnras, 484, 1724B
\bibitem[Chakrabarti (1989)]{c89} Chakrabarti, S. K. 1989, \apj, 347, 365
\bibitem[Chakrabarti \& Titarchuk (1995)]{ct95} Chakrabarti, S. K., \&
Titarchuk, L. 1995, \apj, 455, 623
\bibitem[Chattopadhyay \& Kumar (2016)]{ck16} Chattopadhyay, I., \& Kumar, R. 2016, \mnras, 459, 3792
\bibitem[Crenshaw \& Kraemer (2012)]{ck12} Crenshaw, D. M., \& Kraemer, S. B. 2012, \apj, 753, 75
\bibitem[Das \etal (2014)]{dcnm14} Das, S., Chattopadhyay, I., Nandi, A., \& Molteni, D. 2014, \mnras, 442, 251
\bibitem[Das \& Sharma (2013)]{ds13} Das, U., \& Sharma, P. 2013, \mnras, 435, 2431
\bibitem[Doeleman \etal (2012)]{detal12} Doeleman, S. S., Fish, V. L., Schenck, D. E., \etal 2012, Sci, 338, 355
\bibitem[Dullemond \& Turolla (1998)]{dt98} Dullemond, C. P., \&  Turolla, R. 1998, \apj, 503, 361 
\bibitem[Esin \etal (1997)]{emn97} Esin, A. A., McClintock, J. E., \& Narayan, R. 1997, \apj, 489, 865
\bibitem[Fender \etal (2004)]{fbg04} Fender, R. P., Belloni, T. M., \& Gallo,
E. 2004, \mnras, 355, 1105
\bibitem[Gallo \etal (2003)]{gfp03} Gallo, E., Fender, R. P., \& Pooley, G. G.
2003, \mnras, 344, 60
\bibitem[Ghasemnezhad \& Samadi (2018)]{gs18} Ghasemnezhad, M., \& Samadi, M. 2018, \apj, 865, 93
\bibitem[Gracia \etal (2003)]{gpkc03} Gracia, J., Peitz, J., Keller, Ch., \& Camenzind, M. 2003, \mnras, 344, 468
\bibitem[Gu \& Lu (2000)]{gl00} Gu, W.-M., Lu, J.-F. 2000, \apjl, 540, 33
\bibitem[Gu \& Lu (2004)]{gl04} Gu, W.-M., \& Lu, J.-F. 2004, ChPhL, 21, 2551 
\bibitem[Gu \etal (2009)]{gxll09} Gu, W. M., Xue, L., Liu, T., Lu, J. F. 2009, \pasj, 61, 1313
\bibitem[Gu (2012)]{g12} Gu, W. M. 2012, \apj, 753, 118
\bibitem[Gu (2015)]{g15} Gu, W. M. 2015, \apj, 799, 71
\bibitem[Habibi \etal (2017)]{has17} Habibi, A., Abbassi, S., Shadmehri, M. 2017, \mnras, 464, 5028
\bibitem[Honma (1996)]{h96} Honma, F 1996, \pasj, 48, 77
\bibitem[Jiang \etal (2017)]{jsd17} Jiang, Y.-F., Stone, J., \& Davis, S. W. 2017, arXiv:1709.02845
\bibitem[Jiao \& Wu (2011)]{JW11} Jiao, C.-L., \& Wu, X.-B. 2011, \apj, 733, 112
\bibitem[Jiao \etal (2015)]{Jetl15} Jiao, C.-L., Mineshige, S., Takeuchi, S., \& Ohsuga, K. 2015, \apj, 806, 93
\bibitem[Junor \etal (1999)]{jbl99} Junor, W., Biretta, J. A., \& Livio, M.
1999, Nature, 401, 891
\bibitem[Kato \& Nakamura (1998)]{kn98} Kato, S., \& Nakamura, K. E. 1998, \pasj, 50, 559
\bibitem[Kato \etal (2008)]{kfm08} Kato, S., Fukue, J., \& Mineshige, S. 2008,
 Black-Hole Accretion Disks: Towards a New Paradigm 
 (Kyoto: Kyoto University Press	)
\bibitem[Kumar \& Chattopadhyay (2013)]{kc13} Kumar, R., \& Chattopadhyay, I. 2013, \mnras, 430, 386
\bibitem[Kumar \etal (2013)]{kscc13} Kumar, R., Singh, C. B., Chattopadhyay,
I., \&  Chakrabarti, S. K. 2013, \mnras, 436, 2864
\bibitem[Kumar \etal (2014)]{kcm14} Kumar, R., Chattopadhyay,
I., \& Mandal, S. 2014, \mnras, 437, 2992
\bibitem[Kumar \& Chattopadhyay (2014)]{kc14} Kumar, R., \& Chattopadhyay, I. 2014, \mnras, 443, 3444
\bibitem[Kumar \& Chattopadhyay (2015)]{kc15} Kumar, R., \& Chattopadhyay, I. 2015, ASInC, 12, 93
\bibitem[Kumar \& Chattopadhyay (2017)]{kc17} Kumar, R., \& Chattopadhyay, I. 2017, \mnras, 469, 4221
\bibitem[Kumar \& Gu (2018)]{kg18} Kumar, R., \& Gu, W. M. 2018, \apj, 860, 114
\bibitem[Li \etal (2013)]{los13} Li, J., Ostriker, J., \& Sunyaev, R. 2013, \apj, 767, 105
\bibitem[Lightman \& Eardley (1974)]{le74} Lightman, A. P., \& Eardley, D. M. 1974, \apjl, 187, 1
\bibitem[Lee \etal (2016)]{lckhr16} Lee, S.-J., Chattopadhyay, I., Kumar, R., Hyung, S., \& Ryu, D.
2016, \apj, 831, 33
\bibitem[Lu \etal (1999)]{lgy99} Lu, J. F., Gu, W. M., \& Yuan, F. 1999, \apj, 523, 340
\bibitem[Lu \etal (2004)]{llg04} Lu, J.-F., Lin, Y.-Q., \& Gu, W.-M. 2004, \apjl, 602, 37
\bibitem[Ma \etal (2019)]{mrlw19} Ma, R.-Y., Roberts, S. R., Li, Y.-P. \& Wang, Q. D. 2019, \mnras, 483, 5614
\bibitem[Manmoto \& Kato (2000)]{mk00} Manmoto, T., \& Kato, S. 2000, \apj, 538, 295
\bibitem[Meyer \& Meyer-Hofmeister (1994)]{mm94} Meyer, F., \& Meyer-Hofmeister, E. 1994, A\&A, 288, 175
\bibitem[Meyer \etal (2000)]{mlm00} Meyer, F., Liu, B. F., \& Meyer-Hofmeister, E. 2000, A\&A, 361, 175
\bibitem[Mirabel \etal (1992)]{mrcpl92} Mirabel, I. F., Rodriguez, L. F., Cordier, B., Paul, J., \& Lebrun, F. 
1992, Nature, 358, 215
\bibitem[Molteni \etal (2001)]{mgv01} Molteni, D., Gerardi, G., \& Valenza, M. A. 2001, \apjl, 551, 77
\bibitem[Mosallanezhad \etal (2016)]{mby16} Mosallanezhad, A., Bu, D.-F., \& Yuan, F. 2016, \mnras, 456, 2877
\bibitem[Narayan \& Yi (1994)]{ny94} Narayan, R., \& Yi, I. 1994, \apj, 428, L13
\bibitem[Narayan \& Yi (1995a)]{ny95a} Narayan, R., \& Yi, I. 1995a, \apj, 444, 231
\bibitem[Narayan \& Yi (1995b)]{ny95b} Narayan, R., \& Yi, I. 1995b, \apj, 452, 710
\bibitem[Narayan \etal (1997)]{nkh97} Narayan, R., Kato, S., \& Honma, F. 1997, \apj, 476, 49
\bibitem[Narayan \etal (2012)]{nspk12} Narayan, R., S{\"A} dowski, A., Penna, R. F., \& Kulkarni, A. K. 
2012, \mnras, 426, 3241
\bibitem[Ohsuga \etal (2005)]{omnm05} Ohsuga, K., Mori, M., Nakamoto, T., \& Mineshige, S. 2005, \apj, 628, 3680
\bibitem[Paczy\'nski \& Wiita (1980)]{pw80} Paczy\'nski, B., \& Wiita, P. J. 1980, \aap, 88, 23
\bibitem[Park \etal (2018)]{pkt18} Park, J., Kam, M., Trippe, S. \etal 2018, \apj, 860, 112
\bibitem[Samadi \& Abbassi (2016)]{sa16} Samadi, M., \& Abbassi, S. 2016 \mnras, 455, 3381
\bibitem[S{\c a}dowski \etal (2013)]{snpz13} S{\c a}dowski, A., Narayan, R., Penna, R., \& Zhu, Y. 2013, \mnras, 436, 3856
\bibitem[Sarkar \& Das (2016)]{sd16} Sarkar, B., \& Das, S. 2016, \mnras, 461, 190
\bibitem[Shakura \& Sunyaev(1973)]{ss73} Shakura, N. L., \& Sunyaev, R. A.
1973, \aap, 24, 337
\bibitem[Wandel \& Liang (1991)]{wl91} Wandel, A., \& Liang, E. P. 1991, \apj, 380, 84
\bibitem[Wang \etal (2013)]{wnm13} Wang, Q. D., Nowak, M. A., Markoff, S. B., \etal 2013, Science, 341, 981
\bibitem[Xie \& Yuan(2008)]{xy08} Xie, F.-G., \& Yuan, F. 2008, \apj, 681, 499
\bibitem[Xu \& Chen(1997)]{xc97} Xu, G. H., \& Chen, X. M. 1997, \apj, 489, L29
\bibitem[Xue \& Wang(2005)]{XW05} Xue, L., \& Wang, J.-C. 2005, \apj, 623, 372
\bibitem[Yang \etal (2014)]{yyob14} Yang, X.-H., Yuan, F., Ohsuga, K., \& Bu, D.-F. 2014, \apj, 780, 79
\bibitem[Yuan \etal (2012a)]{ywb12} Yuan, F., Wu, M., \& Bu, D. 2012, \apj, 761, 129
\bibitem[Yuan \etal (2012b)]{ybw12} Yuan, F., Bu, D., \& Wu, M. 2012, \apj, 761, 130
\bibitem[Yuan \etal (2015)]{ygnsbb15} Yuan, F., Gan, Z., Narayan, R., Sadowski, A., Bu, D.-F., \& Bai, X.-N. 2015, \apj, 804, 101
\bibitem[Zeraatgari \etal (2018)]{zmay18} Zeraatgari, F. Z., Mosallanezhad, A., Abbassi, S., \& Yuan, Y.-F. 2018, \apj, 852, 124
\end{thebibliography}
\end{document}